\documentclass[prd,eqsecnum,twocolumn,amsfonts,showpacs]{revtex4}

\input epsf

\usepackage{graphicx}

\usepackage{bm}

\newcommand{\beq}{\begin{equation}}
\newcommand{\eeq}{\end{equation}}
\newcommand{\beqs}{\begin{eqnarray}}
\newcommand{\eeqs}{\end{eqnarray}}

\begin{document}

\title{Partition Function Zeros of a Restricted Potts Model on Self-Dual Strips
of the Square Lattice}

\author{Shu-Chiuan Chang$^{a}$ and Robert Shrock$^{b}$}

\bigskip

\affiliation{(a) \ Department of Physics \\
National Cheng Kung University \\
Tainan 70101, Taiwan} 

\bigskip

\affiliation{(b) \ C. N. Yang Institute for Theoretical Physics \\
State University of New York \\
Stony Brook, N. Y. 11794 }

\begin{abstract}

We calculate the partition function $Z(G,Q,v)$ of the $Q$-state Potts model
exactly for self-dual cyclic square-lattice strips of various widths $L_y$ and
arbitrarily great lengths $L_x$, with $Q$ and $v$ restricted to satisfy the
relation $Q=v^2$.  From these calculations, in the limit $L_x \to \infty$, we
determine the continuous accumulation locus ${\cal B}$ of the partition
function zeros in the $v$ and $Q$ planes.  A number of interesting features of
this locus are discussed and a conjecture is given for properties applicable
for arbitrarily great width.  Relations with the loci ${\cal B}$ for general
$Q$ and $v$ are analyzed.


\end{abstract}

\maketitle

\newpage
\pagestyle{plain}
\pagenumbering{arabic}

\section{Introduction}

The $Q$-state Potts model has served as a valuable model for the study of phase
transitions and critical phenomena \cite{wurev}-\cite{martinbook}.  On a
lattice or, more generally, on a graph $G$ with $n$ sites (vertices), at
temperature $T$, this model is defined by the partition function
\beq
Z(G,Q,v) = \sum_{ \{ \sigma \} }
\exp ( K\sum_{\langle i j \rangle} \delta_{\sigma_i \sigma_j} )
\label{z}
\eeq
where $\sigma_i=1,...,q$ are the classical spin variables on each vertex 
$i \in G$, $\langle i j \rangle$ denotes pairs of adjacent vertices, $K=\beta
J$ where $\beta = (k_BT)^{-1}$, and $J$ is the spin-spin coupling.  We define
\beq
a=e^K, \quad v=a-1 
\label{av}
\eeq
so that $v$ has the physical range of values $0 \le v \le \infty$ and $-1 \le v
\le 0$ for the ferromagnetic and antiferromagnetic cases $J > 0$ and $J < 0$,
respectively.  The (reduced) free energy per site is $f=-\beta F= \lim_{n \to
\infty} n^{-1} \ln Z(G,Q,v)$.  Aside from the $Q=2$ Ising special case on
two-dimensional lattices, the free energy of the model has never been computed
exactly for general $Q$ and $v$ on lattices of dimensionality $d \ge 2$,
although the critical behavior is understood for $d=2$.  In view of this lack
of exact solutions for $d \ge 2$, there is continuing value in studying the
case of lattice strips of finite width $L_y$ and arbitrarily great length
$L_x$, with various boundary conditions.  Although these are
quasi-one-dimensional systems, and the resultant free energy does not exhibit a
phase transition at any finite temperature, they are nevertheless useful for
the exact results that can be obtained and the insights that one can gain
concerning the properties of the model as a function of $Q$ and $v$.  Among
these exact results are the partition function zeros and the loci that they
form in the limit $n \to \infty$.  These loci are determined by the equality in
magnitude of the eigenvalues of the transfer matrix with maximal modulus and
hence are also called the set of equimodular curves.  (Here ``curves'' is used
is a generic sense; these loci can also include line segments.)  In previous
works we have presented exact calculations of the Potts model partition
function for arbitrary $Q$ and $v$ on such strips and have also investigated
various special cases \cite{pref}.  One type of special case is provided
by considering requiring $Q$ and $v$ to satisfy some functional relation.

In this paper we consider the Potts model on self-dual cyclic square-lattice
strips of various widths $L_y$ and arbitrarily great lengths $L_x$, with the
variables $Q$ and $v$ restricted to satisfy, in terms of the variable 
$\zeta_{sq} = v/\sqrt{Q}$, the relation 
\beq
\zeta_{sq}=1 \ . 
\label{zetasq}
\eeq
It will actually be convenient to consider the square of this condition,
namely, 
\beq
Q = v^2 \ . 
\label{qvsq}
\eeq
The condition (\ref{zetasq}) is the condition for the phase transition
temperature of the Potts ferromagnet on a two-dimensional square lattice, i.e.,
$k_B T_c = J/\ln(1+\sqrt{Q} \ )$. (The transition is second-order for $0 < Q
\le 4$ and first-order for $Q > 4$.)  For the $L_x \to \infty$ limits of these
strips we exactly determine the continuous accumulation loci ${\cal B}$ of the
partition function zeros in the $v$ and $Q$ planes. 

Because of the quasi-one-dimensional nature of the infinite-length strips that
we consider, the significance of the condition (\ref{zetasq}) or (\ref{qvsq})
is not that the Potts ferromagnet has a phase transition at this temperature,
but rather that it is a self-dual point; the partition function satisfies the
duality relation
\beq
Z(G,Q,v) \propto Z(G^*,Q,v_d) \ , 
\label{zdual}
\eeq
where $G^*$ denotes the planar dual to $G$ and
\beq
v_d = \frac{Q}{v} \ , \quad i.e., \quad \zeta_d = \frac{1}{\zeta} \ .
\label{vd}
\eeq
If $G=G^*$, as is the case with our self-dual square-lattice strips, and if
$v=v_d$, then, up to a multiplicative factor which will not be important here,
the partition function is transformed to itself under duality.  The properties
of the Potts model for the submanifold of values of $Q$ and $v$ that satisfy
the condition (\ref{qvsq}) are thus of general interest even if the model does
not exhibit a phase transition at the corresponding temperature.

\section{Preliminaries} 

In this section we review some background which will be used in the
subsequent portion of the text.  A graph $G=G(V,E)$ is defined by its vertex
set $V$ and its edge set $E$ (where $V$ should not be confused with the
temperature-like Boltzmann variable $v$).  The number $n$ of vertices of $G$ is
denoted as $n=n(G)=|V|$ and the number of bonds (edges) of $G$ as $e(G)=|E|$.
The degree of a vertex is defined as the number of edges that connect to
it. The Potts model can be generalized from non-negative integer $Q$ to
non-negative real $Q$ via the relation \cite{kf}
\beq
Z(G,Q,v) = \sum_{G^\prime \subseteq G} Q^{k(G^\prime)}v^{e(G^\prime)}
\label{cluster}
\eeq
where $G^\prime$ is a spanning subgraph of $G$, i.e., $G'=(V,E^\prime)$ with
$E^\prime \subseteq E$, and $k(G^\prime)$ denotes the number of connected
components of $G^\prime$.  This relation also allows one to consider $v$ and
$Q$ to be complex, as will be necessary in analyzing the zeros of $Z(G,Q,v)$.
A transfer matrix formalism for the Potts model that can be used for real $Q$
was presented in Ref. \cite{blote82}.

Given that one restricts $Q$ and $v$ to satisfy eq. (\ref{qvsq}), one can
express $Z(G,Q,v)$ as a function of one variable.  Expressing $Q$ as a function
of $v$ via eq. eq. (\ref{qvsq}), one obtains $Z(G,v^2,v)$.  This has the
general form, from eq. (\ref{cluster}), 
\beq
Z(G,v^2,v) = \sum_{G^\prime \subseteq G} v^{2k(G^\prime)+e(G^\prime)}
\label{clusterqvrel}
\eeq
The alternate procedure, expressing $v$ as a function of $Q$, is not as
convenient, since the formal solution for $v$ involves a sign ambiguity, $v =
\pm \sqrt{Q}$.  Clearly, since $Z(G,v^2,v)$ is a polynomial in $v$ with real
coefficients \cite{coeff}, the set of its zeros and the resultant locus ${\cal
B}$ are invariant under complex conjugation.

The Potts model partition function is equivalent to an important function in
mathematical graph theory, the Tutte polynomial $T(G,x,y)$ \cite{wt1,wt2}:
\beq
Z(G,Q,v) = (x-1)^{k(G)}(y-1)^{n(G)} \, T(G,x,y)
\label{zt}
\eeq
where
\beq
x=1+\frac{Q}{v} \ , \quad y=1+v \ .
\label{xyqv}
\eeq
(We follow standard mathematical notation here; the reader should not confuse
these Tutte variables $x$ and $y$ with the longitudinal and transverse strip
graph coordinates $x$ and $y$, since the meaning will be clear from context.)
For a planar graph $G$, the Tutte polynomial satisfies $T(G,x,y) =
T(G^*,y,x)$. In terms of Tutte variable, eq. (\ref{qvsq}) reads
$(x-1)(y-1)=(y-1)^2$.  Since the case $y=1$, i.e., $K=0$, is trivial, one may
concentrate on the case $y \ne 1$ and divide by $(y-1)$, obtaining the
corresponding condition $x=y$.  For a self-dual planar graph
$G$ such as the lattice graphs that we consider here, $T(G,x,y)=T(G,y,x)$.

The Potts model on a $L_x \times L_y$ section of a lattice is related to
quantum algebras, in particular, to the algebra defined by a set of operators
$U_i$ satisfying the relations \cite{tl} 
\beq
U_i^2 = Q^{1/2} U_i
\label{tl1}
\eeq
\beq
[U_i,U_j]=0 \quad {\rm if} \quad |i-j| \ne 1
\label{tl2}
\eeq
\beq
U_i U_{i\pm1} U_i = U_i 
\label{tl3}
\eeq
where the range of the index $i$ on $U_i$ can be taken to be $1 \le i \le
2L_y-1$.  One commonly writes
\beq
\sqrt{Q} = q + q^{-1} 
\label{Qqrel}
\eeq
where 
\beq 
q = e^{i\theta/2}
\label{qdef}
\eeq
so that
\beq
Q=4\cos^2(\theta/2) \ . 
\label{qrel}
\eeq
The case where $q$ is a root of unity, i.e., $\theta/\pi$ is rational, will be
relevant to our discussion below.  We define the root of unity $q_r = e^{i
\pi/r}$ and correspondingly
\beq
Q_r = 4\cos^2(\pi/r) \ . 
\label{qr}
\eeq
The values $Q_r$ corresponding to the roots of unity $q_r$ are often called
Tutte-Beraha numbers; early interest in these arose in studies of the chromatic
polynomial $P(G,Q) = Z(G,Q,-1)$, equal to the Potts antiferromagnet at zero
temperature \cite{wt1,wt2,bkw} (see also \cite{saleur}).  They also play a
special role for the finite-temperature Potts model (with either sign of $J$),
as is clear from the fact that the representations of the algebra of eqs. 
(\ref{tl1})-(\ref{tl3}) are reducible at $Q=Q_r$ \cite{martinber,root1}. We
also define
\beq
v_r = -2\cos(\pi/r)
\label{vr}
\eeq
so that $Q_r = v_r^2$.  For $Q=Q_r$ and $v=v_r$, the associated values of the
Tutte variables are given by eq. (\ref{xyqv}) as $x_r=y_r=1-2\cos(\pi/r)$.
Calculations of zeros of $Z(G,v^2,v)$ on finite sections of the square lattice
found that a subset of these zeros occurred at or near certain $Q_r$ values
\cite{kc}, and, as will be discussed below, we find that the accumulation loci
${\cal B}_Q$ cross the real $Q$ axis at certain $Q_r$ values.

\section{General Form of $Z(G,Q,v)$ for Self-Dual Cyclic Square-Lattice 
Strips}

We focus on strip graphs of the square lattice of length $L_x$ vertices and
width $L_y$ vertices with periodic longitudinal boundary conditions and with
one additional external vertex connected to each of the vertices on one side of
the strip, say the top side.  For these graphs the number of vertices is
$n=L_xL_y+1$, which is equal to the number of faces, and the number of edges is
$e=2n$ \cite{mult}.  The average vertex degree, defined as $2e(G)/n(G)$, is
equal to 4, the same value for the square lattice. As regards individual
vertices, the external vertex has degree $L_x$, the $L_x$ vertices on the
bottom side of the strip have degree 3, and the $L_x(L_y-1)$ other vertices
have degree 4.  These graphs are self-dual, and accordingly we label the
boundary conditions as self-dual cyclic boundary conditions (CDBC) and the
graphs as $G_{CDBC}[L_y \times L_x]$.  The fact that these square-lattice strip
graphs maintain the property of self-duality that holds for the (infinite)
square lattice makes them particularly valuable to use in this type of study.
Furthermore, the periodic longitudinal boundary conditions minimize boundary
effects in this direction.  To explain one of the most important consequences
of these self-dual boundary conditions, we recall that the locus ${\cal
B}_\zeta$ for the Potts ferromagnet on infinite-length, finite-width strips
usually extends to $\zeta=\infty$, i.e., $K=\infty$, reflecting the fact that
the model has a critical point at $T=0$ (and zero magnetization for $T > 0$ but
nonzero magnetization at $T=0$).  This was, indeed, the case for strips of many
different lattices for which we obtained exact determinations of the locus
${\cal B}_\zeta$ or the equivalent loci ${\cal B}_v$ or ${\cal B}_a$ (e.g.,
\cite{a}-\cite{ts}).  However, as our exact determinations of ${\cal B}_\zeta$
for infinite-length, finite-width strips of the square lattice with cyclic or
free self-dual boundary conditions in Ref. \cite{sdg} showed, the locus ${\cal
B}_\zeta$ does not extend to $\zeta=\infty$ and is compact in the $\zeta$
plane.  This difference arises because, owing to the self-duality, it follows
that
\beq
{\cal B}_\zeta \ \ {\rm is \ invariant \ under} \quad \zeta \to \frac{1}{\zeta}
\ . 
\label{binv}
\eeq
Consequently, for the self-dual strips, ${\cal B}_\zeta$ passes through the
zero-temperature point $\zeta=\infty$ if and only if it passes through the
infinite-temperature point, $\zeta=0$.  But it cannot pass through the latter
point because the free energy is analytic in the neighborhood of $K=0$.  Thus,
for general $Q$ and $v$, the infinite-length, finite-width self-dual strips
share with the (infinite) square lattice both the property of self-duality and
the property that ${\cal B}_\zeta$ is compact in the $\zeta$ plane. As we shall
show, however, when we restrict $Q$ and $v$ to obey the relation (\ref{qvsq}),
the resultant singular locus ${\cal B}_v$, and hence also ${\cal B}_Q$, are
noncompact in their respective $v$ and $Q$ planes.  We note that finite
sections of the square lattice with these CDBC boundary conditions have
previously been used for studies of complex-temperature zeros of the Potts
model partition function in Refs. \cite{chw,pfef,kc}.

The Potts model partition function has the general form \cite{dg,sdg}
\beqs
& &  Z(G_{CDBC}[L_y \times L_x],Q,v) = \cr\cr
& = &  Q \, v^{L_xL_y} \, 
\sum_{d=1}^{L_y+1} \ \sum_{j=1}^{n_Z(L_y,d)} \bar\kappa^{(d)}
(\bar \lambda_{L_y,d,j})^{L_x}
\label{zgsum}
\eeqs
where
\beq
\kappa^{(d)} \equiv Q\bar\kappa^{(d)} =
 \sum_{j=0}^{d-1} (-1)^j { 2d-1-j \choose j} Q^{d-j}
\label{kappa}
\eeq
and
\beq
n_Z(L_y,d)=\frac{2d}{L_y+d+1}{2L_y+1 \choose L_y-d+1} 
\label{nzLydGd}
\eeq
for $d=1,...,L_y+1$ (and zero for other values of $d$). The
$\bar\lambda_{L_y,d,j}$ are eigenvalues of a transfer matrix of the type
discussed in Ref. \cite{blote82}.  In terms of the notation of Ref. \cite{sdg},
\beq
\bar\lambda_{L_y,d,j} \equiv v^{-L_y} \lambda_{L_y,d,j} \ . 
\label{lambbar}
\eeq
Since $n_Z(L_y,L_y+1)=1$, i.e., there is only one eigenvalue for $d=L_y+1$, we
label it simply as $\bar\lambda_{L_y,L_y+1,1} \equiv \bar\lambda_{L_y,L_y+1}$.
This eigenvalue is \cite{dg,sdg}
\beq
\bar\lambda_{L_y,L_y+1}=1 \ . 
\label{lamlast}
\eeq
 From the general form (\ref{zgsum}) it follows that $Z(G_{CDBC}[L_y \times
L_x],v^2,v)$ has a zero of multiplicity at least $L_xL_y+2=n+1$ at $v=0$. We
find that this is an isolated zero for these types of graphs. 

The coefficients $\kappa^{(d)}$ can be written as \cite{dg,cf} 
\beq
\kappa^{(d)} =  \prod_{k=1}^d (Q - s_{d,k})
\label{kappafactors}
\eeq
where
\beq
s_{d,k} = 4\cos^2 \bigg ( \frac{\pi k}{2d} \bigg ) \ . 
\label{skd}
\eeq
The first few coefficients $\bar\kappa^{(d)}$ are
$\bar\kappa^{(1)}=1$, $\bar\kappa^{(2)}=Q-2$, and
$\bar\kappa^{(3)}=(Q-1)(Q-3)$.  From eq. (\ref{kappa}) or (\ref{kappafactors})
it follows that if $d \ge 3$ and $d=0$ mod 3, 
$\bar\kappa^{(d)}$ contains the factor $(Q-1)(Q-3)$. i.e., 
\beq
\bar\kappa^{(d)} = (Q-1)(Q-3) \, \mu(Q) \quad {\rm if} \ \ d \ge 3, \ \ 
d=0 \ {\rm mod} \ 3
\label{kappad0mod3}
\eeq
where $\mu(Q)$ is a polynomial of degree $d-3$. This property will be used 
later.  We also note that if $d$ is even, then $\bar\kappa^{(d)}$ contains the
factor $(Q-2)$.  The total number of eigenvalues
is
\beq
\sum_{d=1}^{L_y+1} n_Z(L_y,d) = { 2L_y+1 \choose L_y+1} \ .
\label{nztot}
\eeq
For given values of $Q$, $v$ and $d$, the eigenvalue with maximum modulus,
i.e., the maximal or dominant eigenvalue, is labelled
$\lambda_{L_y,d,max}$.

As discussed in (section 2.2 of) Ref. \cite{a}, the formal definition for the
reduced free energy, $f$ is not adequate for (real, non-negative) non-integer
$Q$ because at certain special points $Q_{sp}$ one has the noncommutativity of
limits
\beq
\lim_{n \to \infty} \lim_{Q \to Q_{sp}} Z(G,Q,v)^{1/n} \ne
\lim_{Q \to Q_{sp}} \lim_{n \to \infty} Z(G,Q,v)^{1/n} \ .
\label{znoncomm}
\eeq
The definitions of $f$ with these two different order of limits are denoted, as
in Ref. \cite{a}, $f_{nQ}$ and $f_{Qn}$.  Clearly, no such issue of
noncommutativity arises if one restricts to positive integer $Q$ values and
uses the original Potts model definition (\ref{z}).  However, since we will
consider the more general case of positive real $Q$ as encompassed in the
formula (\ref{cluster}), we will have to take account of this noncommutativity.

\section{Partition Function Zeros and Loci ${\cal B}$ }

\subsection{General} 

We now determine the loci ${\cal B}$ in the $v$ and $Q$ planes, denoted ${\cal
B}_v$ and ${\cal B}_Q$, for the infinite-length limits of these strip graphs
for several widths $L_y$.  We also compare these to zeros calculated for long
finite strips.  Our procedure is to calculate $Z(G_{CDBC}[L_y \times
L_x],v^2,v)$, solve for the locus ${\cal B}_v$ as the locus of solutions of
degeneracy in magnitude of dominant $\bar\lambda_{L_y,d,j}$'s, and then obtain
${\cal B}_Q$ via the mapping (\ref{qvsq}).  We find that these loci exhibit a
number of interesting systematic properties.  To explain these, we first
discuss our exact solutions for strips of specific widths.
 
\subsection{$L_y=1$} 

In this case, the graph is equivalent to the wheel graph in mathematical graph
theory.  Using our exact solution for the Potts model partition function on
this graph \cite{jz,dg,sdg}, we find that, with $L_x=m$,
\beqs
& & Z(G_{CDBC}[1 \times m],v^2,v) = \cr\cr
& = & v^{m+2} \biggl [ 
\{ (\bar\lambda_{1,1,1})^m + (\bar\lambda_{1,1,2})^m \} + \bar\kappa^{(2)} 
\biggr ]
\label{zsdg1}
\eeqs
where
\beq
\bar\lambda_{1,1,j} = \frac{1}{2}\biggl [ 3+2v \pm \sqrt{5+4v} \, \biggr ]
\label{lam11j}
\eeq
with $j=1,2$ corresponding to $\pm$, and we have used $\bar\lambda_{1,2}=1$
from eq. (\ref{lamlast}).  Here and below the coefficient functions
$\bar\kappa^{(d)}=\bar\kappa^{(d)}(Q)$ are understood to be evaluated on the
manifold (\ref{qvsq}).

In the limit $L_x \to \infty$, we have calculated the exact locus ${\cal B}_v$
in the $v$ plane and its image under the mapping (\ref{qvsq}), ${\cal B}_Q$ in
the $Q$ plane. We find that ${\cal B}_v$ consists of the union of (i) a pinched
oval that crosses the real $v$ axis at $v=v_3=-1$ and $v=-2$ with (ii) a
semi-infinite line segment on the negative real $v$ axis extending leftward
from $v=-2$ to $v=-\infty$.  In the region outside of the pinched oval,
$\bar\lambda_{1,1,1}$ is the dominant eigenvalue.  Along the semi-infinite line
segment, $\bar \lambda_{1,1,1}=\bar \lambda_{1,1,2}^*$, so that $|\bar
\lambda_{1,1,1}|=|\bar \lambda_{1,1,2}|$.  In the region within the oval,
$\bar\lambda_{1,2}=1$ is the dominant eigenvalue.  The fact that ${\cal B}_v$
does not intersect the positive real axis at any finite point is a general
result reflecting the quasi-one-dimensional nature of these infinite-length
strip graphs and the consequent property that the free energy is analytic for
all finite temperatures.  As one moves leftward along the real $v$ axis, the
first nonanalyticity that one encounters, at $v=-1$ is precisely the
zero-temperature limit of the Potts antiferromagnet (with $Q=v^2=1$).

\begin{figure}[hbtp]
\centering
\leavevmode
\epsfxsize=2.4in
\begin{center}
\leavevmode
\epsffile{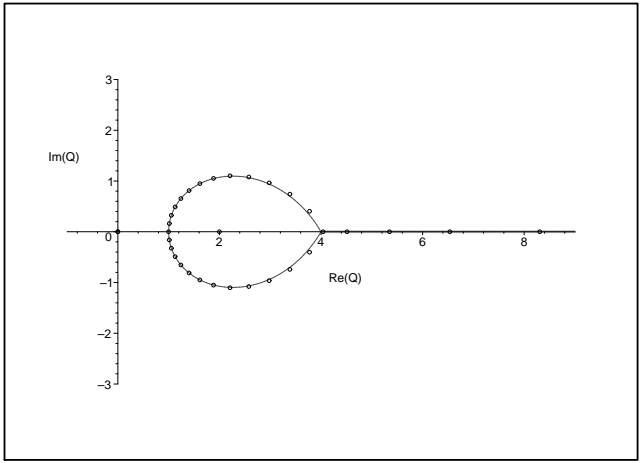}
\end{center}
\vspace{-10mm}
\caption{\footnotesize{Locus ${\cal B}_Q$ for the Potts model on a $1 \times
\infty$ graph with CDBC and $Q=v^2$. This locus includes a semi-infinite line
segment on the interval $4 \le Q \le \infty$. Partition function zeros are
shown for a graph of length $L_x=40$.}}
\label{sdg1q}
\end{figure}

The locus ${\cal B}_Q$, shown in Fig. \ref{sdg1q}, is the union of (i) a
pinched oval that crosses the real $Q$ axis at $Q=Q_3=1$ and $Q=Q_1=4$ (where
$Q_r$ was defined in eq. (\ref{qr})) with (ii) a semi-infinite line segment on
the real $Q$ axis extending from $Q=4$ to $Q=\infty$.
Note that in terms of the Tutte variables,
\beq
v=-2, \ Q=4 \ \Longleftrightarrow \ x = y = -1 \ . 
\label{q4tutte}
\eeq
In Fig. \ref{sdg1q} we also show partition function zeros calculated for a long
finite strip, with $L_x=40$.  (For this length some of the zeros on the
positive real have magnitudes too large to be included in the figure.)  At
$v=-2$ and its image point $Q=4$, the magnitudes of all three eigenvalues are
degenerate and equal to 1.  In addition to the isolated zero at $Q=Q_2=0$ as
discussed above, the partition function has a zero very close to $Q=Q_4=2$.
This can be understood as a consequence of the fact that in this region
$\bar\lambda_{1,2,max}$ is dominant, but its coefficient $\bar\kappa^{(2)}=Q-2$
vanishes at $Q=2$.  In Fig. \ref{sdg1q} one can also see has several real zeros
in the interval $Q > 4$; as $L_x \to \infty$, these and other real zeros in
this interval merge to form the semi-infinite line segment on ${\cal B}_Q$.

This calculation provides a simple example of the noncommutativity
(\ref{znoncomm}).  Consider the point $v_{sp} = -\sqrt{2}$, whence $Q_{sp}
\equiv v_{sp}^2 = 2$.  Now if we first set $Q=2$ and then take $n \to \infty$
to calculate $f_{nQ}$, then, since the coefficient $\bar\kappa^{(2)}$ vanishes
at this point, the dominant term $\bar\lambda_{1,2}^m=1$ in eq.  (\ref{zsdg1})
is annihilated.  Furthermore, at this point 
\beq
\bar\lambda_{1,1,j} = (\sqrt{2}-1)e^{\pm i\theta}
\label{lampolar}
\eeq
where
\beq
\theta = {\rm arctan} \biggl ( \frac{\sqrt{4\sqrt{2}-5}}{3-2\sqrt{2}} \biggr )
\label{theta}
\eeq 
so that 
\beq
Z(G_{CDBC}[1 \times m],2,-\sqrt{2}) = 2(-2)^{\frac{m+2}{2}}(\sqrt{2}-1)^m
\cos(m\theta) 
\label{zspec}
\eeq
and hence 
\beq
\lim_{n \to \infty} \lim_{(Q,v) \to (2,-\sqrt{2})} |Z^{1/n}| = 2-\sqrt{2}
\label{lim1}
\eeq
With the opposite order of operations, one obtains the different result 
\beq
\lim_{(Q,v) \to (2,-\sqrt{2})} \lim_{n \to \infty}|Z^{1/n}| = \sqrt{2} 
\label{znq}
\eeq
where, since there is no canonical phase to pick for the $1/n$'th root
\cite{a}, we just display the modulus.

\subsection{$L_y=2$}

\begin{figure}[hbtp]
\centering
\leavevmode
\epsfxsize=2.4in
\begin{center}
\leavevmode
\epsffile{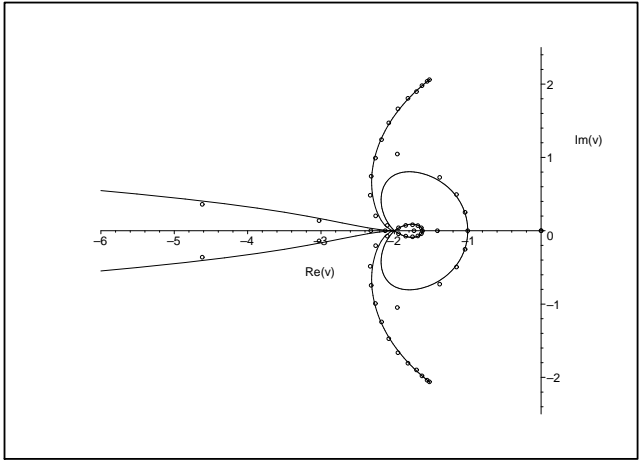}
\end{center}
\vspace{-10mm}
\caption{\footnotesize{Locus ${\cal B}_v$ for the Potts model on a $2 \times
\infty$ graph with CDBC and $Q=v^2$.  Partition function zeros are shown for a
graph of length $L_x=30$.}}
\label{sdg2v}
\end{figure}

\begin{figure}[hbtp]
\centering
\leavevmode
\epsfxsize=2.4in
\begin{center}
\leavevmode
\epsffile{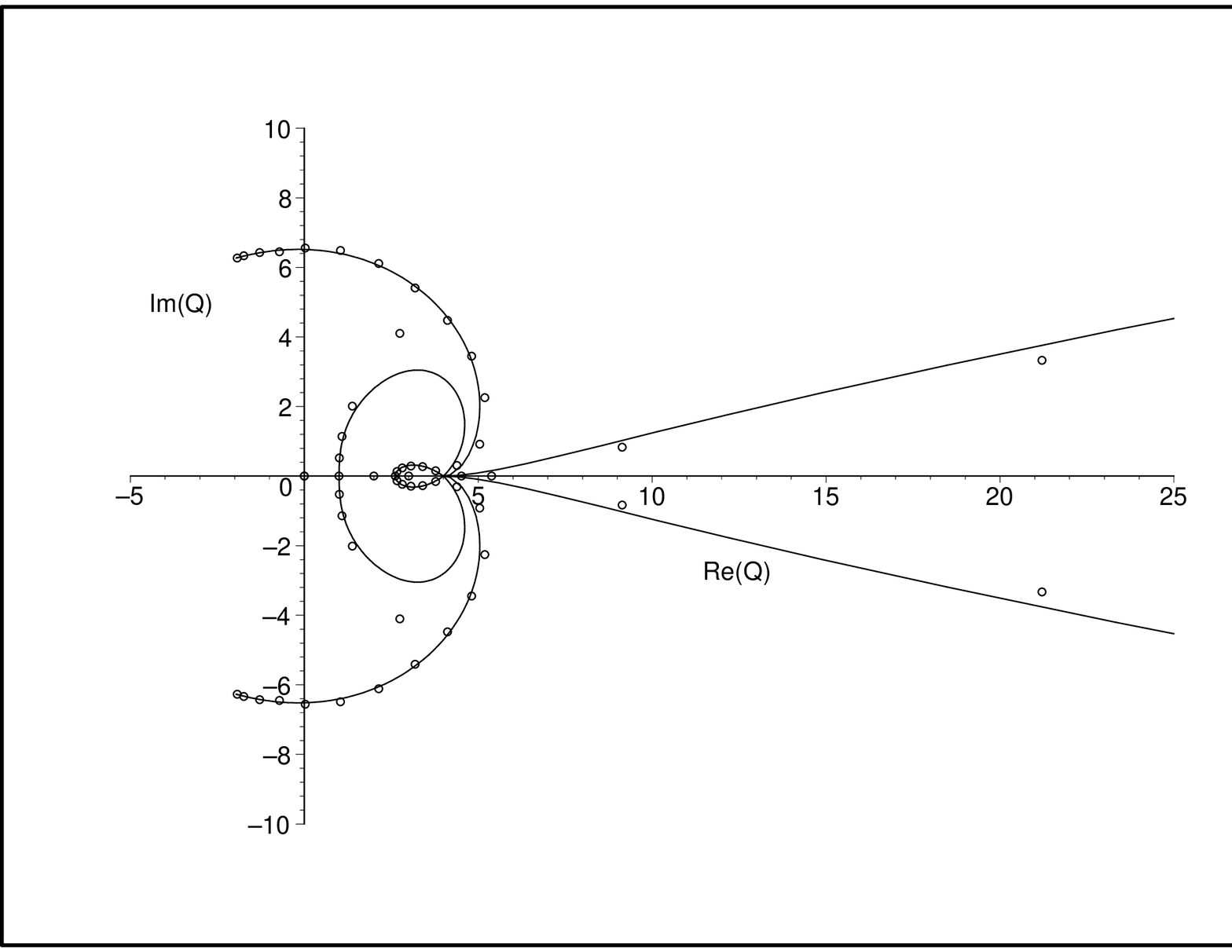}
\end{center}
\vspace{-10mm}
\caption{\footnotesize{Locus ${\cal B}_Q$ for the Potts model on a $2 \times
\infty$ graph with CDBC and $Q=v^2$.  Partition function zeros are shown for a
graph of length $L_x=30$.}}
\label{sdg2q}
\end{figure}

\begin{figure}[hbtp]
\centering
\leavevmode
\epsfxsize=2.4in
\begin{center}
\leavevmode
\epsffile{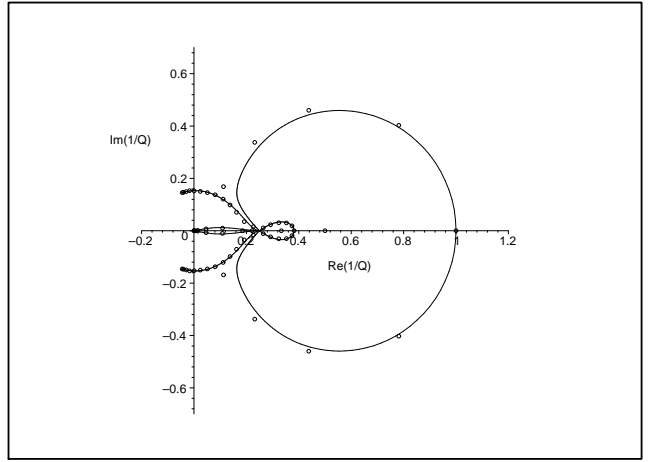}
\end{center}
\vspace{-10mm}
\caption{\footnotesize{Locus ${\cal B}$ plotted in the $1/Q$ plane, for the
Potts model on a $2 \times \infty$ graph with CDBC and $Q=v^2$.  Partition
function zeros are shown for a graph of length $L_x=30$.}}
\label{sdg2iq}
\end{figure}

For the width $L_y=2$ strips with CDBC we use our previous exact calculation of
the partition function in Refs. \cite{dg,sdg} and specialize to the submanifold
of variables given by (\ref{qvsq}). From eq. (\ref{nzLydGd}) there are five
$\bar\lambda_{2,1,j}$'s, and four $\bar\lambda_{2,2,j}$'s, so that, together
with the single $\lambda_{2,3}$, the total number of $\bar\lambda_{2,d,j}$'s is
ten.  We have
\beqs
& & Z(G_{CDBC}[2 \times m],v^2,v) = v^{2m+2} \, \biggl [ 
\sum_{j=1}^5 (\bar\lambda_{2,1,j})^m \cr\cr
& + & 
\bar\kappa^{(2)} \sum_{j=1}^4 (\bar\lambda_{2,2,j})^m 
+ \bar\kappa^{(3)} \biggr ]
\label{zsdg2}
\eeqs
where
\beq
\bar \lambda_{2,2,j}=\frac{1}{4}\biggl [ 7+\eta \sqrt{5} \, + 4v \pm  
\Bigl [2(19+7\eta \sqrt{5} \, ) + 8(3 +\eta\sqrt{5} \, )v \Bigr ]^{1/2} 
\ \biggr ]
\label{lam22j}
\eeq
where $\eta= \pm 1$, yielding four possible terms. The $\bar \lambda_{2,1,j}$
are roots of a quintic equation which can be derived from the general quintic
given in Ref. \cite{sdg}.  For $v=-2$, all of the roots (\ref{lam22j}) have
unit modulus and this is also true of the roots of the above-mentioned quintic
equation, so that all ten $\bar \lambda_{2,d,j}$ have $|\bar
\lambda_{2,d,j}|=1$.  We show our results for ${\cal B}_v$ and ${\cal B}_Q$ in
Figs. \ref{sdg2v} and \ref{sdg2q}, together with partition function zeros
calculated for a strip with length $L_x=30$.  (For this and other figures, some
zeros have sufficiently large magnitudes so that they are not shown.)  We find
that the locus ${\cal B}_v$ crosses the negative $v$ plane at $v=-2$ and
$v=-2\cos(\pi/(2\ell+1))$ for $\ell=1,2$, i.e., $-1$, and $-2\cos(\pi/5) =
-(1+\sqrt{5})/2 \simeq -1.618$.  Correspondingly, ${\cal B}_Q$ crosses the
positive $Q$ axis at the squares of these values, $4\cos^2(\pi/(2\ell+1))$ for
$\ell=0,1,2$, i.e., 4, 1, and $4\cos^2(\pi/5) = (3+\sqrt{5})/2 \simeq
2.618$. Two complex-conjugate curves on ${\cal B}$ extend to complex infinity
in the $v$ and $Q$ planes, rendering the loci noncompact in these variables.
In contrast, ${\cal B}$ is compact in the $1/v$ or $1/Q$ planes, and we show
${\cal B}$ in the $1/Q$ plane in Fig. \ref{sdg2iq}.  The fact that ${\cal B}$
extends to $1/Q=0$ means that the free energy does not have a Taylor series
expansion in $1/Q$.  This feature is similar to the situation for the quantity
$W = \lim_{n \to \infty} P(G,Q)^{1/n}$ for certain families of graphs
\cite{qinf}.  The confluence of curves at $v=-2$ and thus $Q=4$ reflects the
degeneracy in magnitude of the ten $\bar\lambda_{2,d,j}$'s.

We comment on some additional features of ${\cal B}_Q$ in Fig. \ref{sdg2q}.
This locus contains a set of complex-conjugate arcs that extend from $Q=4$
outward from the real axis, crossing the imaginary axis at $Q \simeq \pm 6.52i$
and ending at $Q \simeq -1.96 \pm 6.27i$.  Along the real $Q$ axis for $Q \ge
4$, the dominant term is $\bar\lambda_{2,1,max}$.  A different
$\bar\lambda_{2,1,max}$ is dominant in the region extending to complex infinity
in the $Q$ plane separated by curves on ${\cal B}_Q$ from this wedge containing
the interval $Q \ge 4$.  In the region containing the real interval $Q_3 \le Q
\le Q_5$, i.e., $1 \le Q \le (1/2)(3+\sqrt{5}) \simeq 2.618$,
$\bar\lambda_{2,2,max}$ is dominant, while in the region containing the real
interval $Q_5 \le Q \le Q_1$, i.e., $(1/2)(3+\sqrt{5}) \le Q \le 4$,
$\bar\lambda_{2,3}=1$ is dominant.  The partition function also has several
isolated zeros that are not on ${\cal B}_Q$.  In addition to the zero at $Q=0$,
it has zeros in the interval $[0,4]$ very near to $Q=Q_4=2$ and $Q=Q_6=3$.  The
zeros very near to $Q_4$ and $Q_6$ result from the fact that the coefficient
$\bar\kappa^{(2)}$ that multiplies the dominant term
$(\bar\lambda_{2,2,max})^m$ in the region around $Q=2$ vanishes at $Q=2$, and
the coefficient $\bar\kappa^{(3)}$ that multiplies the dominant term
$(\bar\lambda_{2,3})^m=1$ in the region around $Q=3$ vanishes at $Q=3$.  The
partition function also has real zeros in the region $Q > 4$.

\subsection{$L_y=3$}

\begin{figure}[hbtp]
\centering
\leavevmode
\epsfxsize=2.4in
\begin{center}
\leavevmode
\epsffile{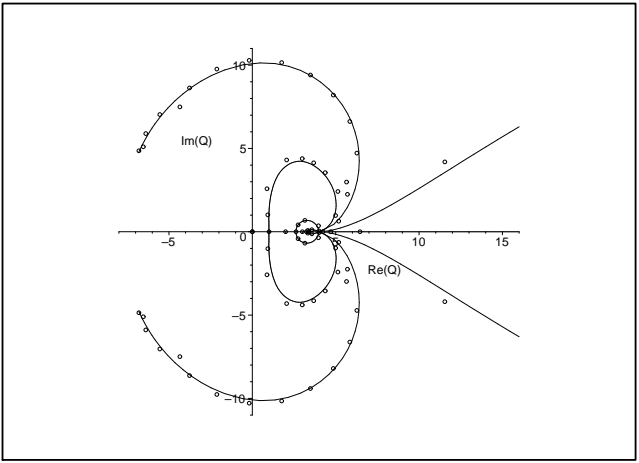}
\end{center}
\vspace{-10mm}
\caption{\footnotesize{Locus ${\cal B}_Q$ for the Potts model on a $3 \times
\infty$ graph with CDBC and $Q=v^2$.  This locus includes a semi-infinite line
segment in the interval $4 \le Q \le \infty$. Partition function zeros are
shown for a graph of length $L_x=26$. }}
\label{sdg3q}
\end{figure}

\begin{figure}[hbtp]
\centering
\leavevmode
\epsfxsize=2.4in
\begin{center}
\leavevmode
\epsffile{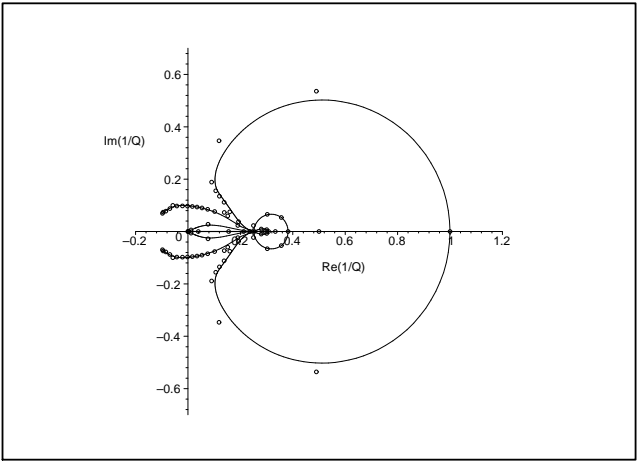}
\end{center}
\vspace{-10mm}
\caption{\footnotesize{Locus ${\cal B}$ plotted in the $1/Q$ plane, for the
Potts model on a $3 \times \infty$ graph with CDBC and $Q=v^2$.  This locus
includes a line segment in the interval $0 \le 1/Q \le 1/4$. Partition
function zeros are shown for a graph of length $L_x=26$.}}
\label{sdg3iq}
\end{figure}

For $L_y=3$, from the exact calculation of $Z(G_{CDBC}[3 \times L_x],Q,v)$
performed for Ref. \cite{sdg} we specialize to the manifold (\ref{qvsq}) and
calculate the partition function zeros.  Because of the large number of
$\bar\lambda_{3,d,j}$'s (a total of 35) that enter, we do not list these
explicitly here.  Having discussed above how the zeros in the $v$ plane and the
locus ${\cal B}_v$ map to zeros and the locus ${\cal B}_Q$ in the $Q$ plane, we
display our results only in the latter plane.  In Fig. \ref{sdg3q} we show a
plot of zeros and ${\cal B}_Q$.  To depict the portions of ${\cal B}_Q$ that
extend infinitely far from the origin, it is again useful to include, as
Fig. \ref{sdg3iq}, a plot of ${\cal B}$ in the $1/Q$ plane, where it is
compact.  We find that in the interval $-2 \le v \le -1$ the locus ${\cal B}_v$
crosses the negative real $v$ axis at $v=-2$ and $v=-2\cos(\pi/(2\ell+1))$ for
$\ell=1,2,3$, i.e., $-1$, $-2\cos(\pi/5) = -(1/2)(1+\sqrt{5}) \simeq -1.618$,
and $-2\cos(\pi/7) \simeq -1.802$.  Correspondingly, in the interval $1 \le Q
\le 4$ the locus ${\cal B}_Q$ crosses the positive $Q$ axis at the squares of
these values, $4\cos^2(\pi/(2\ell+1))$ for $\ell=0,1,2,3$, i.e., 4, 1,
$4\cos^2(\pi/5) = (3+\sqrt{5})/2 \simeq 2.618$, and $4\cos^2(\pi/7) \simeq
3.247$.  For $L_x=26$, there are zeros very close to $Q_4=2$, $Q_6=3$, and
$Q_8=2+\sqrt{2} \simeq 3.414$.  Our exact results also show that the locus
${\cal B}_v$ contains a semi-infinite line segment on the real axis extending
leftward from $v=-2$ to $v=\infty$ and, correspondingly, ${\cal B}_Q$ contains
a semi-infinite line segment on the interval $4 \le Q \le \infty$.  While the
region boundary that includes $Q=1$ and $Q=4$ was convex in the neighborhood of
$Q=1$ for $L_y=1$ and $L_y=2$, it becomes concave in this neighborhood for
$L_y=3$ (and $L_y=4$).  This boundary also extends further outward
from the real axis than for the lower values of $L_y$, reaching to roughly $\pm
4.23i$.  The outermost finite complex-conjugate arcs reach farther away
from the real axis than the analogous arcs for $L_y=2$, crossing the imaginary
axis at $Q \simeq \pm 10i$, and extend farther to the left, with arc endpoints
at approximately $Q \simeq -6.8 \pm 4.8i$.  The dominant terms are as follows:
(i) two different $\bar\lambda_{3,1,max}$'s in the region containing the real
interval $Q \ge 4$ and the separate region extending to complex infinity, (ii)
$\bar\lambda_{3,2,max}$ in the region containing the interval $Q_3 \le Q \le
Q_5$, (iii) $\bar\lambda_{3,3,max}$ in the region containing the interval $Q_5
\le Q \le Q_7$, i.e., approximately $2.618 < Q < 3.247$, and (iv)
$\bar\lambda_{3,4}=1$ in the region containing the interval $Q_7 \le Q \le
Q_1=4$.

\subsection{$L_y=4$}

For the present work we have also calculated $Z(G_{CDBC}[4 \times L_x],Q,v)$
exactly for $Q=v^2$.  Owing to the large number of $\bar\lambda_{4,d,j}$'s (126
in total), we again do not list them here.  We find that the resultant locus
${\cal B}_v$ crosses the negative real $v$ axis at $v=-2$ and
$v=-2\cos(\pi/(2\ell+1))$ for $\ell=1,2,3,4$ and thus ${\cal B}_Q$ crosses the
positive $Q$ axis at the squares of these values.  The outer complex-conjugate
arcs on ${\cal B}_Q$ continue the trend observed for $L_y=2,3$ of having parts
that extend farther from the real axis, crossing the imaginary axis at $Q
\simeq \pm 13i$, but at the same time, having endpoints that move leftward and
inward toward the real axis, so that these endpoints occur near to this axis.
The dominant terms are as follows: (i) different $\bar\lambda_{4,1,max}$'s in
the region containing the interval $Q \ge 4$ and the region extending to
complex infinity, (ii) $\bar\lambda_{4,2,max}$ in the region containing the
interval $Q_3 \le Q \le Q_5$, (iii) $\bar\lambda_{4,3,max}$ in the region
containing the interval $Q_5 \le Q \le Q_7$, (iv) $\bar\lambda_{4,4,max}$ in
the region containing the interval $Q_7 \le Q \le Q_9$, and (v)
$\bar\lambda_{4,5,}=1$ in the region containing the interval $Q_9 \le Q \le
Q_1=4$.  In addition to the isolated zero at $Q=Q_2=0$, the partition function
also exhibits other isolated zeros not on ${\cal B}_Q$ very close to $Q_4=2$,
$Q_6=3$, $Q_8 =2+\sqrt{2} \simeq 3.414$, and $Q_{10}=(1/2)(5+\sqrt{5}) \simeq
3.618$.  The reason is again due to vanishing coefficients of dominant terms at
these values of $Q$, as discussed above.

\section{General Properties and Conjecture for ${\cal B}$ for Arbitrary 
$L_y$} 

In our exact calculations on CDBC strips with widths $1 \le L_y \le 4$, we have
discerned several systematic features which we generalize here.  For the Potts
model on self-dual cyclic strips of the square lattice of arbitrary width
$L_y$, with $Q$ and $v$ satisfying eq. (\ref{qvsq}), in the limit $L_x \to
\infty$, we conjecture that 
\begin{itemize}

\item 
In the interval $-2 \le v \le -1$ the locus ${\cal B}_v$ crosses the negative
real $v$ axis at $v=-2$ and
\beqs
v & = & v_{2\ell+1} = -2\cos \left ( \frac{\pi}{2\ell+1} \right ) \cr\cr
& & \quad {\rm for} \quad 1 \le \ell \le L_y
\label{vcross}
\eeqs
For odd $L_y$ this locus ${\cal B}_v$ also includes the semi-infinite line
segment
\beq
-\infty \le v \le -2 \ . 
\label{vline}
\eeq

\item 
Hence, in the interval $1 \le Q \le 4$ the locus ${\cal B}_Q$ crosses the $Q$
axis at
\beqs
Q & = & Q_{2\ell+1} =4\cos^2 \left ( \frac{\pi}{2\ell+1} \right ) \cr\cr
& & \quad {\rm for} \quad 0 \le \ell \le L_y
\label{qcross}
\eeqs
corresponding to the roots of unity given in eq. (\ref{qr}) with $r=2\ell+1$.
For odd $L_y$ the locus ${\cal B}_Q$ also includes the semi-infinite line
segment
\beq
4 \le Q \le \infty \ . 
\label{qline}
\eeq
These properties do not preclude the possibility that, for sufficiently wide
strips, the locus ${\cal B}_Q$ may also cross the real $Q$ axis at points in
the intervals $Q < 0$ or $Q > 4$.

\item 
For a strip of width $L_y$, the locus ${\cal B}_v$ divides the $v$ plane into
at least $2L_y$ regions, and the locus ${\cal B}_Q$ divides the $Q$ plane into
at least $2L_y$ regions.  (Note that the number of regions into which ${\cal
B}_v$ and ${\cal B}_Q$ divide their respective planes need not be equal; for a
gedanken example, if ${\cal B}_v$ were to contain just two complex-conjugate
arcs emanating from $v=-2$ and crossing the imaginary $v$ axis at $\pm i b_0$
with $b_0 \ne 0$ and then terminating in endpoints, this locus would not
enclose a region in the $v$ plane, but its image in the $Q$ plane would enclose
a region.)

\item

In the respective regions of the $v$ and $Q$ planes away from the real axis and
extending to complex infinity, the dominant eigenvalue is
$\bar\lambda_{L_y,1,max}$.  In the region of the $v$ plane containing the
interval $-\infty \le v \le -2$, and equivalently, in the region of the $Q$
plane containing the interval $4 \le Q \le \infty$, the dominant eigenvalue is
a different $\bar\lambda_{L_y,1,max}$.  The fact that this is different is
evident from the equimodular curve separating this region in the $v$ plane from
the region extending to complex infinity, and similarly in the $Q$ plane.  For
odd $L_y$, two eigenvalues from this set becomes degenerate on the real axis in
the respective intervals $-\infty \le v \le -2$ and $4 \le Q \le \infty$,
giving rise to the line segments on ${\cal B}_v$ and ${\cal B}_Q$.

\item 

Concerning the other regions that contain intervals of the real $v$ and $Q$
axes, the $L_y=1$ case has been solved exactly above: in the region $-2 \le v
\le -1$ and the corresponding region $1 \le Q \le 4$, $\bar\lambda_{1,2}=1$ is
dominant.  For $L_y \ge 2$, in the region of the $v$ plane containing the
interval $v_{2k+3} \le v \le v_{2k+1}$, or equivalently in the region in the
$Q$ plane containing the interval $Q_{2k+1} \le Q \le Q_{2k+3}$ for $1 \le k
\le L_y-1$, $\bar\lambda_{L_y,k+1,max}$ is dominant. In the respective regions
of the $v$ and $Q$ planes containing the intervals $-2 \le v \le v_{2L_y+1}$
and $Q_{2L_y+1} \le Q \le Q_1=4$, the dominant eigenvalue is
$\bar\lambda_{L_y,L_y+1}=1$.

\item 

All of the $\bar\lambda_{L_y,d,j}$'s have the magnitude
$|\bar\lambda_{L_y,d,j}|=1$ at $v=-2$ or, in terms of $Q$, at $Q=4$.
Hence, all of the curves on ${\cal B}_v$ meet at the point $v=-2$,
and correspondingly, all of the curves on ${\cal B}_Q$ meet at the point $Q=4$.

\item 

We next remark on zeros of $Z(G_{CDBC}[L_y \times L_x],v^2,v)$ that, in the
limit $L_x \to \infty$, do not lie on ${\cal B}$, but instead are isolated
(discrete).  We have shown (so this is not just a conjecture) that there is an
isolated zero at $v=0$ of multiplicity at least $n+1$ and hence also at $Q=0$
with multiplicity that scales like $n/2$ for large $n$.  In addition, from our
exact results, we conjecture that for arbitrary $L_y$, $Z(G_{CDBC}[L_y \times
L_x],Q,v)$ with $Q=v^2$ has zeros close to the values
\beq
Q = Q_{2\ell} \quad {\rm for} \quad 1 \le \ell \le L_y+1 \ . 
\label{qcrossnear}
\eeq
The basis of this inference and the way to understand this result is that that
these points are located within the region where the dominant eigenvalue of the
transfer matrix, $\bar\lambda_{L_y,d,max}$, has a coefficient
$\bar\kappa^{(d)}$ in eq. (\ref{zgsum}) that vanishes at the points
(\ref{qcrossnear}), as is evident from eqs. (\ref{kappafactors}) and
(\ref{skd}).

\end{itemize}

For each of our plots showing the exact loci ${\cal B}_Q$ on infinite-length
strips of the square lattice with cyclic self-dual boundary conditions and
$Q=v^2$, we have also shown zeros calculated for long finite strips for
comparison. As was evident, these zeros lie close to the asymptotic loci ${\cal
B}$.  This comparison is especially easy to make for the plots of ${\cal B}$ in
the $1/Q$ plane, where it is compact; clearly, in the $Q$ plane, the zeros for
long finite sections cannot have arbitrarly large modulus and hence cannot
track the portions of ${\cal B}$ that extend infinitely far from the origin.
It is also of interest to compare our results with calculations of zeros of
$Z(G_{CDBC}[L \times L],Q,v)$ for finite $L \times L$ sections of the square
lattice with $Q=v^2$ in Ref. \cite{kc}.  Of course, for such a finite section,
no locus ${\cal B}_Q$ is defined; one is only able to make a rough comparison
of patterns of zeros.  One sees that a number of zeros in the $Q$ plane occur
at or near to certain $Q_r$'s \cite{kc} and that the zeros in the $v$ and
$Q$ planes exhibit patterns suggesting the importance of the points $v=-2$ and
$Q=4$.

For general $Q$ and $v$, the locus ${\cal B}_\zeta$ is compact for the
infinite-length, finite-width self-dual strips of the square lattice
\cite{sdg}.  We noted above how these strips share this property, along with
their self-duality, with the (infinite) square lattice.  In contrast, with
these strips, when one restricts $Q$ and $v$ to satisfy the relation
(\ref{qvsq}), ${\cal B}_v$ and ${\cal B}_Q$ become noncompact in the respective
$v$ and $Q$ planes, passing through $1/v=0$ and $1/Q=0$.

\section{Self-Dual Strips with Free Longitudinal Boundary Conditions}

We have also calculated $Z(G,v^2,v)$ on self-dual strips of the square lattice
with free longitudinal boundary conditions.  A strip of this type of length
$L_x$ and width $L_y$ is constructed by adding an external vertex which is
connected to each of the vertices on one horizontal side, say the upper one,
and to each of the vertices on one vertical side, say the right-hand one, of
the strip \cite{chw,pfef,sdg}. (Hence, the upper right-most vertex of the strip
is connected to the external vertex with a double edge.)  We denote this type
of boundary condition as ``free (self)-dual boundary condition'', abbreviated
as FDBC, and the graph as $G_{FDBC}[L_y \times L_x]$.

For the strip $G_{FDBC}[1 \times m]$ we use our previous exact calculation of
$Z$ \cite{dg,sdg} and restrict $Q$ and $v$ to satisfy (\ref{qvsq}).  The
partition function depends only on (the $m$'th powers of) the
$\bar\lambda_{1,1,j}$ in eq. (\ref{lam11j}).  As is evident from
eq. (\ref{lam11j}), these are equal in magnitude for real $v \le -5/4$, and
consequently, ${\cal B}_v$ is the semi-infinite line segment extending from
$v=-5/4$ to $v=-\infty$.  The image of this in the $Q$ plane, ${\cal B}_Q$, is
the semi-infinite line segment from $Q=(5/4)^2=1.5625$ to $Q=\infty$.  Thus, as
was the case with the infinite-length CDBC strips, the loci ${\cal B}_v$ and
${\cal B}_Q$ are noncompact in the respective $v$ and $Q$ planes.  For FDBC
strips with larger widths we find that ${\cal B}_v$ and ${\cal B}_Q$ consist of
arcs and line segments and generically exhibit the above noncompactness.

\section{Other Strips}

We have considered strips of the square lattice with free ($F$) and periodic
($P$) boundary conditions in the transverse and longitudinal directions, which
we label in an obvious manner, e.g., $(FBC_y,FBC_x)=$ free, $(FBC_y, PBC_x)=$
cyclic, $(PBC_y, FBC_x)=$ cylindrical, and $(PBC_y, PBC_x)=$ toroidal.  We
observe that for the cyclic square-lattice strips that we have studied, ${\cal
B}_Q$ crosses the real axis at a set of points including $Q_1=4$ and
$Q_{2\ell}$ for $1 \le \ell \le L_y$.  We find that for some strips ${\cal
B}_Q$ also crosses the real axis for $Q < 0$ or $Q > 4$.  

In addition, we have calculated the loci ${\cal B}$ for infinite-length strips
of the triangular lattice, with $Q$ and $v$ restricted to satisfy the relation
$Q=v^2(v+3)$, which is the phase transition condition for the Potts
ferromagnetic on the triangular lattice.  As in the case of the cyclic and
toroidal square-lattice strips, we find that for the case of periodic boundary
longitudinal conditions, ${\cal B}_Q$ crosses the real axis at $Q=0$ and $Q=4$
as well as at other points that depend on the type of strip. The intersections
of ${\cal B}_Q$ with the real axis sometimes include finite line segments.  A
detailed discussion of our results for these other lattice strips will be
presented elsewhere.

\section{Relation to ${\cal B}$ for General $Q$, $v$}

It is interesting to observe the connections between the loci that we have
determined for the $Q=v^2$ case to the continuous accumulation loci ${\cal B}$
in other variables, such as (i) complex $v$ for fixed $Q$; (ii) complex $Q$ for
fixed $v$; and (iii) complex $\zeta$ for, e.g., various values of $Q$. We
concentrate on the CDBC strips and comment briefly on others. In
Ref. \cite{sdg} we displayed the loci in the $\zeta$ plane for several values
of $Q$ in Ref.  \cite{sdg}.  In relating these results to our present ones, one
must take account of the noncommutativity (\ref{znoncomm}).  Let us consider,
for example, the CDBC strip of width $L_y=1$. Figs. 9, 11, 12, and 13 of that
paper gave ${\cal B}_\zeta$ for $Q=2$, 4, 5, and 100, calculated by first
setting $Q$ equal to the given value, and then taking $L_x \to \infty$; these
plots show that this locus passes through $\zeta=-1$, and hence our special
case $Q=v^2$.  This agrees with our present finding that for this special case,
${\cal B}_Q$ passes through $Q=Q_1=4$, and contains a semi-infinite line
segment on the interval $4 \le Q \le \infty$.  In the case $Q=2$, the fact that
the locus was calculated in Ref. \cite{sdg} by first setting $Q$ equal to this
value and then taking $L_x \to \infty$ explains why the locus ${\cal B}_\zeta$
in Ref. \cite{dg} passes through $\zeta=-1$, while our locus ${\cal B}_Q$,
calculated by first taking $L_x \to \infty$, does not pass through $Q=2$; the
reason is that these limits do not commute.  The origin of this
noncommutativity is clear: setting $Q=2$ first means that $\bar\kappa^{(2)}=0$,
and this annihilates what would be the dominant term in the vicinity of
$v=-\sqrt{2}$, $Q=2$, namely, $(\bar\lambda_{1,2})^{L_x}=1$.  For $Q=3$, the
locus ${\cal B}_\zeta$, shown in Fig. 10 of Ref. \cite{sdg} does not pass
through $\zeta=-1$, which again agrees with our finding that the point $Q=3$ is
not on ${\cal B}_Q$ for the special case $Q=v^2$.  Similar comparisons can be
made for the results shown in Ref. \cite{sdg} for CDBC strips with widths
$L_y=2$ and $L_y=3$.

The case of self-dual strips of the square lattice with free longitudinal
boundary conditions was also studied in Ref. \cite{sdg}. For example, as is
shown in Figs. 2-6 of Ref. \cite{sdg}, as $Q$ increases from 1 to $(5/4)^2$,
two complex-conjugate arcs on ${\cal B}_\zeta$ enlongate and two of their
endpoints come together and pinch the real axis at $\zeta=-1$; for $Q \ge
(5/4)^2$, the locus ${\cal B}_\zeta$ continues to pass through $\zeta=-1$.
Again, this agrees with our present analysis, where we have found that for
$Q=v^2$, ${\cal B}_v$ and ${\cal B}_Q$ contain the respective semi-infinite
line segments $-\infty \le v \le -5/4$ and $(5/4)^2 \le Q \le \infty$.  Similar
connections can be made for other strips.

We can also relate our present results to the loci ${\cal B}_Q$ that we
previously determined in the $Q$ plane for the $L_x \to \infty$ limits of
chromatic polynomials $P(G,Q)=Z(G,Q,-1)$ of various lattice strip graphs with
CDBC and FDBC \cite{dg}.  For CDBC strips we found that ${\cal B}$ crossed the
real $Q$ axis at $Q=Q_3=1$, which lies on the submanifold $Q=v^2$ and agrees
with our present result that for $Q$ and $v$ restricted to this submanifold,
${\cal B}_v$ passes through $v=-1$ (see Figs. 2-4 of Ref. \cite{dg}).  For
strips of the square lattice with cyclic and toroidal (and M\"obius and
Klein-bottle) boundary conditions, we found that for $v=-1$, ${\cal B}_Q$ does
not pass through $Q=1$.  This agrees with our finding that for these strips,
with $Q=v^2$, ${\cal B}_v$ and ${\cal B}_Q$ do not pass through the respective
points $v=-1$ and $Q=1$.

\section{Conclusions}

In conclusion, we have presented exact results for the continuous accumulation
set ${\cal B}$ of the locus of zeros of the Potts model partition function for
the infinite-length limits of self-dual cyclic square-lattice strips of various
widths $L_y$ with $Q$ and $v$ satisfying the relation $Q=v^2$.  For these
quasi-one-dimensional strips, one motivation for interest in this submanifold
of values of $Q$ and $v$ is its invariance under a duality transformation. From
our exact calculations we have discerned several general features and have
incorporated them in a conjecture applicable for arbitrary width $L_y$.  We
have also presented some results on ${\cal B}$ for self-dual strips of the
square lattice with free longitudinal boundary conditions.  Further studies of
these loci for other strips are worthwhile.  The findings add to the set of
exact results that one has for the Potts model on the $n \to \infty$ limits of
lattice graphs.

\section{Acknowledgments}

This research was partially supported by the NSF grant
PHY-00-98527 (R.S.) and the Taiwan NSC grant NSC-94-2112-M-006-013 (S.-C.C.).

\vfill
\eject


\begin{thebibliography}{99}

\bibitem{wurev}
F. Y. Wu, Rev. Mod. Phys. {\bf 54}, 235 (1982). 

\bibitem{baxter}
R. J. Baxter, {\it Exactly Solved Models in Statistical Mechanics} (Academic
Press, New York, 1982). 

\bibitem{martinbook}
P. Martin, {\it Potts Models and Related Problems
in Statistical Mechanics} (World Scientific: Singapore, 1991). 

\bibitem{pref} 
References to some of our earlier works and other related papers
in this area can be found in S.-C. Chang and R. Shrock, Physica A {\bf 346},
400 (2005); {\it ibid.} {\bf 347}, 314 (2005).  We switch notation here
relative to our previous papers, using $Q$ for what was denoted $q$ there in
order to follow conventional notation for $q$ as a root of unity. 

\bibitem{kf}
P. Kasteleyn and C. Fortuin, J. Phys. Soc. Jpn. {\bf 26} (Suppl.) 11 (1969);
C. Fortuin and P. Kasteleyn, Physica {\bf 57}, 536 (1972). 

\bibitem{blote82}
H. W. J. Bl\"ote and M. P. Nightingale, Physica A {\bf 112}, 405 (1982).

\bibitem{coeff} 
The coefficients are, of course, not just real, but are
non-negative integers; however, the latter property is not needed for the
invariance noted here. 

\bibitem{wt1}
W. T. Tutte,  J. Combin. Theory {\bf 9}, 289 (1970).

\bibitem{wt2}
W. T. Tutte, ``Chromials'', in Lecture Notes in Mathematics, v. 411, p. 
243 (1974); W. T. Tutte, {\it Graph Theory}, vol. 21 of Encyclopedia of
Mathematics and Applications (Addison-Wesley, Menlo Park, 1984).

\bibitem{tl}
H. N. V. Temperley and E. H. Lieb, Proc. R. Soc. A {\bf 322}, 251 (1971).

\bibitem{bkw}
S. Beraha, J. Kahane, and N. Weiss, J. Combin. Theory B {\bf 27}, 1 (1979); 
J. Combin. Theory B {\bf 28}, 52 (1980).

\bibitem{saleur}
H. Saleur, Commun. Math. Phys.{\bf 132}, 657 (1990); Nucl. Phys. B {\bf
360}, 219 (1991).

\bibitem{martinber}
P. Martin, J. Phys. A {\bf 20}, L399 (1987). 

\bibitem{root1} Indeed, more generally, representations of quantum algebras
have special properties when the deformation parameter, analogous to $q$ here,
is a root of unity.  Reviews include G. Lusztig, {\it Introduction to Quantum
Groups}, Prog. Math.  {\bf 110} (Birkh\"auser, Boston, 1993); C. Kassel, {\it
Quantum Groups} (Springer, New York, 1995); and V. Chari and A. Pressley, {\it
Guide to Quantum Groups} (Cambridge Univ. Press, Cambridge, 1994).

\bibitem{mult} 
Note that for $L_x=2$, the graph has two-fold multiple horizontal
edges joining adjacent vertices. 

\bibitem{chw}
C. N. Chen, C. K. Hu, and F. Y. Wu, Phys. Rev. Lett. {\bf 76}, 169 (1996). 

\bibitem{pfef}
V. Matveev and R. Shrock, Phys. Rev. E {\bf 54}, 6174 (1996).
In this paper, the CDBC and FDBC strips were denoted DBC2 and DBC1. 

\bibitem{luwu}
W. T. Lu and F. Y. Wu, Physica A {\bf 258}, 157 (1998). 

\bibitem{kc}
S.-Y. Kim and R. Creswick, Phys. Rev. E {\bf 63}, 066107 (2001).

\bibitem{dg}
S.-C. Chang and R. Shrock, Physica A {\bf 301}, 301 (2001). 

\bibitem{sdg} 
S.-C. Chang and R. Shrock, Phys. Rev. E {\bf 64}, 066116 (2001). 

\bibitem{jz}
S.-C. Chang and R. Shrock, Physica A {\bf 301}, 196 (2001). 

\bibitem{cf}
S.-C. Chang and R. Shrock, Physica A {\bf 296}, 131 (2001).

\bibitem{a}
R. Shrock, Physica A {\bf 283}, 388 (2000). 

\bibitem{s3a}
S.-C. Chang and R. Shrock, Physica A {\bf 296}, 234 (2001).

\bibitem{ta}
S.-C. Chang and R. Shrock, Physica A {\bf 286}, 189 (2000).

\bibitem{hca}
S.-C. Chang and R. Shrock, Physica A {\bf 296}, 183 (2001).

\bibitem{ka}
S.-C. Chang and R. Shrock, Int. J. Mod. Phys. B {\bf 15}, 443 (2001).

\bibitem{ts} 
S.-C. Chang, J. Salas, and R. Shrock, J. Stat. Phys. {\bf 107} 1207 (2002).

\bibitem{qinf}
R. C. Read and G. F. Royle, in {\it Graph Theory, Combinatorics, and
Applications} (Wiley, New York, 1991), v. 2, p. 1009; 
R. Shrock and S.-H. Tsai, Phys. Rev. E {\bf 56}, 3935 (1997); 
J. Phys. A {\bf 31}, 9641 (1998); A. Sokal, Combin. Prob. Comput. {\bf 10}, 
41 (2001).

\end{thebibliography}
\end{document}